\documentclass[12pt]{article}
\begin{document}
\begin{center}
{\bf Deformed Quantum Field Theory, Thermodynamics at Low and High
Energies, and Gravity. II}\\
\vspace{5mm} A.E.Shalyt-Margolin \footnote{E-mail:
a.shalyt@mail.ru; alexm@hep.by}\\ \vspace{5mm} \textit{National
Center of Particles and High Energy Physics, Bogdanovich Str. 153,
Minsk 220040, Belarus}
\end{center}
\vspace{0.5cm}
\begin{abstract}
This work is a continuation of studies presented in the papers
arXiv: 0911.5597, 1003.4523. In the work it is demonstrated that
with the use of one and the same parameter deformation may be
described for several cases of the General Relativity within  the
scope of both the Generalized Uncertainty Principle (UV-cutoff)
and the Extended Uncertainty Principle (IR-cutoff). All these
cases have a common thermodynamic interpretation of the
corresponding gravitational equations. Consideration is given to
the possibility for extension of the obtained results to more
general cases. Possible generalization of the uncertainty relation
for the pair (cosmological constant, "space-time volume"), where
the cosmological constant is regarded as a dynamic quantity at
high and low energies is analyzed.
\end{abstract}

\section{Introduction}
In the last decade numerous works devoted to a Quantum Field
Theory (QFT) at Planck's scale \cite{Planck1}--\cite{Planck3} have
been published(of course, the author has no pretensions of being
exhaustive in his references). This interest stems from the facts
that (i) at these scales it is expected to reveal the effects of a
Quantum Gravity (QG), and this still unresolved theory is
intriguing all the researchers engaged in  the field of
theoretical physics; (ii) modern accelerators, in particular LHC,
have the capacity of achieving the energies at which some QG
effects may be exhibited.
\\ Now it is clear that ÷òî a Quantum Field Theory (QFT) at Planck's scales,
and possibly at very large scales as well, undergoes changes
associated with the appearance of additional parameters related to
(i) a minimal length (on the order of the Planck's length)and
(ii)a minimum momentum. As this takes place, the corresponding
parameters are naturally considered as deformation parameters,
i.e. the related quantum theories are considered as a high-energy
deformation (at Planck's scales) and a low-energy deformation
(IR-cutoff), respectively, of the well-known quantum field theory,
the latter being introduced in the corresponding high- and
low-energy limits and exact to a high level.   The deformation is
understood as an extension of a particular theory by inclusion of
one or several additional parameters in such a way that the
initial theory appears in the limiting transition \cite{Fadd}.
\\Most natural approach to the introduction of the above-mentioned parameters
is to treat a quantum field theory with the Generalized
Uncertainty Principle (GUP) \cite{Ven1}--\cite{Kim1} and with the
Extended Uncertainty Principle (EUP), respectively
\cite{Bole:05}--\cite{Kim1}. In the case of GUP we easily obtain a
minimal length on the order of the Planck's $l_{min}\sim l_{p}$
and the corresponding high-energy deformation of well-known
QFT--QFT with GUP. It should be noted that QFT with GUP at
Planck's scales (Early Universe) is attested in many works (for
example \cite{Ven1}--\cite{Magg1})/ Even if we disregard the works
devoted to a string theory, still remaining a tentative one, GUP
is quite naturally derived from the gedanken experiment
\cite{GUPg1}--\cite{Ahl1}.
\\ On the other hand, GUP has no way in the spaces with large
length scales (for example (A)dS). For such spaces, e.g., in \cite
{Bole:05},\cite{Park} the Extended Uncertainty Principle has been
introduced (find its exact definition below) giving an absolute
minimum in the uncertainty of the momentum.
\\ The problem is to find whether there are cases when
the deformations generated by GUP and EUP are defined by the same
parameter. By author's opinion this is the case for Gravity
modified (deformed) within GUP and EUP, when the corresponding
initial theory has a "thermodynamic interpretation"
\cite{Jac1}--\cite{Cai1}. Specifically, the deformation parameter
$\alpha=l_{min}^{2}/x^{2}$,$l_{min}\sim l_{p}$,$0<\alpha\leq1/4$
where $x$ is the measuring scale, introduced by the author in a
series of works \cite{shalyt1}--\cite{shalyt13} meets the above
requirements.
\\ Note that this parameter has been introduced to study
the deformation of QFT at Planck's scale, although the deformation
per se, associated with a high-energy modification of the density
matrix, was "minimal" in that it presented no noncommutativity
operators related to different spatial coordinates
\begin{equation}\label{UDPl}
[X_{i},X_{j}]\neq 0, i\neq j
\end{equation}
and hence "limited" as in the end it failed to lead to GUP.
Nevertheless, the corresponding deformation parameter in some way
is universal.
\\This paper continues the studies, described in \cite{shalyt-aip}
--\cite{shalyt-gravity} (the latter in particular), of the
fundamental quantities in "thermodynamic interpretation" of
gravity \cite{Jac1}--\cite{Cai1} for GUP and EUP deformations of
the latter. Compared to the works \cite{shalyt-gravity}, the
results from which are used in this paper, the important results
associated with EUP are put forward together with the
demonstration that GUP and EUP  have the same deformation
parameter, at least in this context.
\\ The structure of this work is as follows.
In Section 2 it is shown that the deformation of the fundamental
thermodynamic quantities for black holes within GUP and EUP may be
interpreted with the use of the same parameter. In Section 3,
within the scope of a dynamic model for the cosmological constant
$\Lambda$ (vacuum energy density), GUP is studied for the pair
($\Lambda,V$) \cite{shalyt-aip},\cite{shalyt-entropy2}, where $V$
-- is the "space-time volume". In this Section consideration is
given to the possible existence of EUP for this pair, i.e. to a
possible extension of the Uncertainty Principle to the pair in the
IR region, and hence to the possible substantiation of the proper
(coincident with the experimental) value for $\Lambda$. In Section
4 the results of Section 2 are applied to Einstein's Equations for
space with horizon and to Friedmann's Equations. It is
demonstrated that in both cases their deformation (in the first
case within GUP and in the second case within EUP) may be
interpreted with the use of the same small dimensionless parameter
having a known variability domain.
\\And, finally, in Section 5 the problems of further investigations are discussed,
some final comments are given.

\section{Universal Deformation Parameter
 in Gravitational Thermodynamics with GUP and EUP}
In this Section the Gravitational Thermodynamics (GT) is
understood as thermodynamics of spaces with horizon
\cite{Padm11},\cite{Padm13}.
\subsection{Gravitational Thermodynamics with GUP}
We use the notation and principal results from \cite{Park}. So, GUP is of the form
\begin{equation}\label{UDP2.1}
 \Delta x_{i}\Delta p_{j}\geq \hbar \delta_{ij} [ 1 + \alpha^{\prime 2}
 l^2_{p}
\frac{(\Delta p_{i})^{2}}{\hbar ^2}]
\end{equation}
and, since $ \Delta x_{i}\Delta p_{j}>\hbar \delta_{ij}$, we have
\begin{equation}\label{UDP2.2}
\Delta x_{i}\Delta p_{i}\geq  \hbar \delta_{ij} [ 1 +
\alpha^{\prime 2}
 \frac{l^2_{p}}{\Delta x_{i}^{2}}
\frac{(\Delta p_{i})^{2}\Delta x_{i}^{2}}{\hbar ^2}]>\hbar
\delta_{ij}[ 1 + \frac{1}{4}\alpha_{\Delta x_{i}}],
\end{equation}
where $\alpha_{\Delta x_{i}}$ -- parameter $\alpha$ corresponding
to $\Delta x_{i}$, $l_{min}=2\alpha^{\prime}l_{p}$. Besides, as
distinct from \cite{Park}, for the dimensionless factor in GUP,
instead of $\alpha$, we use $\alpha^{\prime}$ to avoid confusion
with the deformation parameter.
\\ In this terms the uncertainty in moment is given
by the nonstrict inequality
\begin{equation}\label{UDP2.4}
2\hbar(\alpha_{\Delta x_{i}}\Delta
x_{i})^{-1}[1-\sqrt{1-\alpha_{\Delta x_{i}}}]\leq\Delta p_{i}\leq
2\hbar(\alpha_{\Delta x_{i}}\Delta
x_{i})^{-1}[1+\sqrt{1-\alpha_{\Delta x_{i}}}].
\end{equation}
But for the quantities determining GT in terms of $\alpha$ one can
derive exact expressions. Indeed, in terms of $\alpha$ the
GUP-modification (or rather GUP-deformation)is easily obtained for
the Hawking temperature
\cite{acs}--\cite{Nou},\cite{Park},\cite{Kim1} that has been
computed in the asymptotically flat  $d$ - dimensional space for a
Schwarzshild black hole with a metric given by
\begin{equation}\label{UDP2.5}
 ds^2=-N^2 dt^2 +N^{-2} dr^3 +r^2 d \Omega^2_{d-2},
\end{equation}
where
\begin{equation}
N^2= 1-\frac{16 \pi G M}{(d-2) \Omega_{d-2} r^{d-3}},
\end{equation}
$\Omega_{d-2}$ is the area of the unit sphere $S^{n-2}$, and
$r_{+}$ is the uncertainty in the emitted particle position by the
Hawking effect, expressed as
\begin{equation}\label{UDP2.6}
\Delta x_{i}\approx r_{+}
\end{equation}
and being nothing else but a radius of the event horizon. In this
case the deformation parameter $\alpha$ arises naturally.
Actually, modification of the Hawking temperature is of the form,
see formula (10) in \cite{Park}
\begin{equation}\label{UDP2.7}
T_{GUP}=(\frac{d-3}{4\pi})\frac{\hbar r_{+}}{2\alpha^{\prime
2}l^{2}_{p}}[1-(1-\frac{4\alpha^{\prime 2}l_{p}^{2}}{r_{+}^{2}})^{1/2}]
\end{equation}
and may be written in a natural way as
\begin{equation}\label{UDP2.71}
T_{GUP}=2(\frac{d-3}{4\pi})\frac{\hbar}{r_{+}} \alpha_{r_{+}}
^{-1} [1-(1-\alpha_{r_{+}})^{1/2}],
\end{equation}
where $\alpha_{r_{+}}$- parameter $\alpha$ associated with
$r_{+}$. It is clear that $T_{GUP}$  is actually the deformation
$T_{Hawk}$ -- black hole temperature for a semiclassical case
\cite{Hawk3}. In such a manner compared to $T_{Hawk}$ $T_{GUP}$ is
additionally dependent only on the dimensionless small deformation
parameter $\alpha_{r_{+}}$.
\\The dependence of the black hole entropy on $\alpha_{r_{+}}$ may be derived
in a similar way. For a semiclassical approximation of the
Bekenstein-Hawking formula \cite{Bek1},\cite{Hawk3}
\begin{equation}\label{UDP2.8}
S=\frac{1}{4}\frac{A}{l^{2}_{p}},
\end{equation}
where $A$ -- surface area of the event horizon, provided the
horizon event is of radius $r_+$, $A\sim r^{2}_+$ and
(\ref{UDP2.8}) is clearly of the form
\begin{equation}\label{UDP2.81}
S=\sigma \alpha^{-1}_{r_{+}},
\end{equation}
where $\sigma$ is some dimensionless denumerable factor. The
general formula for quantum corrections \cite{mv} given as
\begin{equation}\label{UDP2.9}
S_{GUP} =\frac{A}{4l_{p}^{2}}-{\pi\alpha^{\prime 2}\over 4}\ln
\left(\frac{A}{4l_{p}^{2}}\right) +\sum_{n=1}^{\infty}c_{n}
\left({A\over 4 l_p^2} \right)^{-n}+ \rm{const}\;,
\end{equation}
where the expansion coefficients $c_n\propto \alpha^{\prime
2(n+1)}$ can always be computed to any desired order of accuracy
\cite{mv}, may be also written in the general case as a Laurent
series in terms of $\alpha_{r_{+}}$
\begin{equation}\label{UDP2.91}
S_{GUP}=\sigma \alpha^{-1}_{r_{+}}-{\pi\alpha^{\prime 2}\over
4}\ln (\sigma \alpha^{-1}_{r_{+}}) +\sum_{n=1}^{\infty}(c_{n}
\sigma^{-n}) \alpha^{n}_{r_{+}}+ \rm{const}.
\end{equation}
In what follows the representation in terms of the deformation
parameter $\alpha$ is referred to as {\bf
$\alpha$-representation}.
\subsection{Gravitational Thermodynamics with EUP}
Let us consider QFT with EUP \cite{Park}. In this case we obtain
QFT with $p_{min}$. Obviously, there is no minimal length
$l_{min}$ in QFT with EUP whatsoever but we assume that QFT with
GUP is valid. At the present time for such an assumption we can
find solid argumentation \cite{GUPg1}--\cite{Ahl1}. As will be
shown later, in this case the fundamental quantities may be also
expressed in terms of $\alpha$. Hereinafter we use {\bf a small
dimensionless parameter}
\begin{equation}\label{EUP01}
\alpha_{\widetilde{l}}=\frac{l^{2}_{or}}{\widetilde{l}^{2}},
\end{equation}
where $l_{or}\equiv l_{original}=2\alpha^{\prime}l_{p}$,
$\alpha^{\prime}$--dimensionless constant on the order of unity from GUP
(\ref{UDP2.1}), and it is suggested that
\begin{equation}\label{EUP02}
2l_{or}\leq \widetilde{l}\quad, i.e \quad
0<\alpha_{\widetilde{l}}\leq 1/4.
\end{equation}
Similar to the previous Section, it is convenient to use the
principal results of \cite{Park} (sections 3,4). Then EUP in (A)dS
space takes the form
\begin{equation}\label{EUP}
 \Delta x_{i}\Delta p_{j}\geq \hbar \delta_{ij} [1 + \beta^{2}
\frac{(\Delta x_{i})^{2}}{\l^2}],
\end{equation}
where $l$ is the characteristic, large length scale $l\gg l_{p}$
and $\beta$ is a dimensionless real constant on the order of unity
\cite{Park}. From EUP there is an absolute minimum in the momentum
uncertainty???
\begin{equation}\label{EUP0}
\Delta p_{i}\geq \frac{2\hbar \beta}{l} ,
\end{equation}
EUP (\ref{EUP}) may be rewritten as
\begin{equation}\label{EUP1}
 \Delta x_{i}\Delta p_{j}\geq \hbar \delta_{ij} [1 + \beta^{2}
\frac{(\Delta x_{i})^{2}}{\l_{or}^2}\frac{\l_{or}^2}{\l^2}]=\hbar
\delta_{ij} [ 1 + \beta^{2} \alpha_{l}\alpha^{-1}_{\Delta x_{i}}].
\end{equation}
Considering that in a theory with fixed $l\gg l_{p}$
\begin{equation}\label{EUP2}
\alpha_{l}=const\ll 1,
\end{equation}
 (\ref{EUP}),(\ref{EUP1}) may be written as
\begin{equation}\label{EUP3}
 \Delta x_{i}\Delta p_{j}\geq \hbar \delta_{ij}
[ 1 + \beta^{2}\alpha_{l} \alpha^{-1}_{\Delta x_{i}}]=\hbar
\delta_{ij} [ 1 + \widetilde{\beta}^{2}\alpha^{-1}_{\Delta
x_{i}}],
\end{equation}
where $\beta$ is redetermined as
\begin{equation}\label{EUP3.1}
\beta\mapsto\widetilde{\beta}=\sqrt{\alpha_{l}}\beta.
\end{equation}
However, in this case $\beta$ may be left as it is, whereas
$\alpha$ may be redetermined because $\alpha^{-1}_{\Delta x_{i}}$
in (\ref{EUP1}),(\ref{EUP3}) is not a small parameter. In
consequence we can redetermine $\alpha$ as
\begin{equation}\label{EUP4}
\widetilde{\alpha}_{\Delta x_{i}}=\alpha_{l} \alpha^{-1}_{\Delta
x_{i}},
\end{equation}
where $\widetilde{\alpha}_{\Delta x_{i}}$ is now a small parameter.
\\Owing to such a {\bf duality}, EUP
(\ref{EUP}),(\ref{EUP1}) may be rewritten in terms of a new small
parameter $\widetilde{\alpha}$ similar to $\alpha$ as follows:
\begin{equation}\label{EUP5}
 \Delta x_{i}\Delta p_{j}\geq \hbar \delta_{ij}
[ 1 + \beta^{2} \widetilde{\alpha}_{\Delta x_{i}}].
\end{equation}
Then in analogy with \cite{Park} (Section 3), for Hawking temperature
of the $d$-dimensional Schwarzshild-AdS black hole with the metric
function we have
\begin{equation}\label{EUP6}
N^2= 1+ \frac{r^2}{l^2_{AdS}}-\frac{16 \pi G M}{(d-2) \Omega_{d-2}
r^{d-3} }
\end{equation}
in the metric of(\ref{UDP2.5}) and the cosmological constant
$\Lambda=-(d-1)(d-2)/2 l^2_{AdS}$.
\\ Therewith the $\alpha$-representation of the Hawking temperature $T_{EUP}$
\cite{Park} (formula (15)) takes the form
\begin{equation}\label{EUP7}
 T_{EUP(AdS)}=(\frac{d-3}{4\pi}){\frac{\hbar }{r^{2}_{+}}}[ 1+
 \frac{(d-1)}{(d-3)}\alpha^{-1}_{r_{+}}\alpha_{l_{AdS}}]=(\frac{d-3}{4\pi}){\frac{\hbar }{r^{2}_{+}}}[ 1+
 \frac{(d-1)}{(d-3)}\widetilde{\alpha}_{r_{+}}].
\end{equation}
In the same way we can easily obtain the $\alpha$-representation of the
Hawking temperature for a Schwarzshild-AdS black hole and for a combined case
 ((formula (28) from the \cite{Park})) of GUP and EUP -- (GEUP)
\begin{eqnarray}\label{EUP9}
T_{GEUP(AdS)}=2(\frac{d-3}{4\pi})\frac{\hbar}{r_{+}}
\alpha_{r_{+}} ^{-1}
[1-\sqrt{1-\alpha_{r_{+}}[1+\frac{\alpha_{l_{AdS}}(d-1)}{(d-3)}\alpha^{-1}_{r_{+}}]}]\nonumber \\
=2(\frac{d-3}{4\pi})\frac{\hbar}{r_{+}} \alpha_{r_{+}} ^{-1}
[1-\sqrt{\frac{d-3-\alpha_{l_{AdS}}(d-1)}{(d-3)}-\alpha_{r_{+}}}],
\end{eqnarray}
i.e. in the general case we get a Laurent series from $\alpha$.
\\Similarly, we can obtain the $\alpha$-representation for the corresponding value of $T_{GEUP(dS)}$ ((formula (32) from
\cite{Park}) in the de Sitter (dS) space by the substitution
$l^{2}_{AdS}\rightarrow -l^{2}_{dS}.$
\\ Note that, as it has been indicated in \cite{shalyt-aip},
\cite{shalyt-entropy2},
 $\alpha_{r_{+}} ^{-1}$   has one more interesting feature
\begin{equation} \label{comm6}
\alpha_{r_{+}} ^{-1}\sim {r_{+}}^{2}/l_{p}^{2}\sim S_{BH}.
\end{equation}
Here $S_{BH}$ is the Bekenstein-Hawking semiclassical black hole
entropy with the characteristic linear size $r_{+}$. For example,
in the spherically symmetric  case $r_{+}=R$ - radius of the
corresponding sphere with the surface area $A$, and
\begin{equation} \label{comm7}
A=4\pi r_{+}^{2},S_{BH}=A/4l_{p}^{2}=\frac{\pi}{4\alpha^{\prime
2}} \alpha_{r_{+}}^{-1}.
\end{equation}
In \cite{Kim1} GUP and EUP are combined by the principle called
the Symmetric Generalized Uncertainty Principle (SGUP):
\begin{equation}\label{SGUP1}
  \Delta x \Delta p \ge \hbar \left( 1 + \frac{(\Delta x)^2}{L^2} +
      \l^2 \frac{(\Delta p)^2}{\hbar^2} \right),
\end{equation}
where $l\ll L$ and $l$ defines the limit of the UV-cutoff (not
being such up to a constant factor as in the case of GUP).Then  a
minimal length is determined as
\begin{center}
 $\Delta x_{\rm min} =
2l/\sqrt{1-4\l^2/L^2}$,
\end{center}
whereas $L$ defines the limit for IR-cutoff i. e. we have a
minimum momentum
\begin{center}
$\Delta p_{\rm min} = 2\hbar/(L\sqrt{1-4\l^2/L^2)}$.
\end{center}
 And using the Euclidian action
formalism by Gibbons and Hawking \cite{Hawk4}, in \cite{Kim1} the
corresponding correction of the Hawking temperature for an
ordinary(not A(dS)) Schwarzshild-black hole is computed. This
correction is given as $T_{SGUP}$. In the notation of this work
\begin{center}
$\Delta x_{\rm min} =
2l/\sqrt{1-4\alpha_{l}^{-1}\alpha_{L}}=2l/\sqrt{1-4\widetilde{\alpha_{l}}}$,
\end{center}
where $\widetilde{\alpha_{l}}$--small parameter introduced in conformity with (\ref{EUP4}).
We can easily obtain the $\alpha$-representation for
$T_{SGUP}$ that is completely similar to the
$\alpha$-representation of $T_{GEUP(AdS)}$.
\\ It should be noted that in the realistic theories $l\sim l_{p}$, and
it is obvious that $(\sqrt{1-4\l^2/L^2)}\approx 1$. Thus, $\Delta
x_{\rm min}\approx 2l \sim 2l_{p}$ and hence in this case we get a
minimal length that is much the same (to within $\alpha^{\prime}$)
as in the case of GUP. It is seen that, with due regard for the
requirement $l\ll L$, $\Delta p_{\rm min}$ is derived close (to
within $\beta$) to $\Delta p_{\rm min}$ (\ref{EUP0}) in a theory
with EUP.
\\
\\ The question arises as to
{\bf what for all these manipulations with
writing and rewriting of the already derived expressions in the
$\alpha$-representation are necessary}.
\begin{center}
\end{center}
{\bf 2.1} Owing to this procedure, we can draw the conclusion that
all the quantities within the scope of the stated problem are
dependent on one and the same deformation parameter $\alpha$ that
is small, dimensionless (discrete in the case of GUP), and varying
over the given interval.  And, provided the infrared cutoff $l$ is
defined, we have
\begin{center}
$\alpha_{l}=\alpha_{min}=l^{2}_{or}/l^{2}\leq\alpha\leq 1/4$ and
$l_{or}\equiv l_{original}\sim l_{p}$.
\end{center}
If we primordially consider a theory with GUP only, then
$l_{or}\equiv l_{min}$. But in the arbitrary case it is required
that $l_{or}=2\alpha^{\prime}l_{p}$, where $\alpha^{\prime}$ is a
certain dimensionless constant on the order of unity.
\\ The property of discreteness is retained for $\alpha$ in the cases when
only GUP (without generalizations)is valid because  in this case
the length seems to be quantized, the lengths being considered
from $2l_{min}$ rather than from $l_{or}=l_{min}$ as a singularity
arises otherwise \cite{shalyt2}--\cite{shalyt9}.
\\
\\{\bf 2.2} Actually, all the quantities may be represented
as a Laurent series in terms of $\alpha$, and a solution of the
problem at hand may be understood as finding of the members in
this series.
\\
\\{\bf 2.3} When the problem has separate solutions for the cases including the UV- and IR-cutoffs,
we can consider expansion in each of the cases in terms of their
own small parameters: $\alpha$ in the case of UV-cutoff and
$\widetilde{\alpha}$ in the case of IR-cutoff, where
$\widetilde{\alpha}$ {\bf is a duality} of $\alpha$
\begin{center}
$\widetilde{\alpha}_{\widetilde{l}}=\alpha_{l}
\alpha^{-1}_{\widetilde{l}},$
\end{center}
$l$ determines, to within a factor on the order of unity, the characteristic system's size, and $l\gg l_{p}$.
\section{The Cosmological
Constant Problem and QFT with GUP and SGUP}
In this section it is assumed that $\Lambda$ may be varying in time.
Generally speaking, $\Lambda$ is referred to as a constant just
because it is such in the equations, where it occurs: Einstein
equations \cite{Einst}. But in the last few years the dominating
point of view has been that $\Lambda$  is actually a dynamic
quantity, now weakly dependent  on time  \cite{Ran1}--\cite{Shap}.
It is assumed therewith that, despite the present-day smallness of
$\Lambda$ or even its equality to zero, nothing points to the fact
that this situation was characteristics for the early Universe as
well. Some recent results \cite{Min1}--\cite{Min4} are rather
important pointing to a potentially dynamic character of
$\Lambda$. Specifically, of great interest is the Uncertainty
Principle derived in these works for the pair of conjugate
variables $(\Lambda,V)$:
\begin{equation}\label{CC1}
\Delta\Lambda\, \Delta V \sim \hbar,
\end{equation}
where $\Lambda$ is the vacuum energy density (cosmological
constant). It is a dynamic value fluctuating around zero; $V$ is
the space-time volume. Here the volume of space-time $V$ results
from the Einstein-Hilbert action $S_{EH}$ \cite{Min2}:
\begin{equation}\label{CC2}
\Lambda \int d^{4}x \sqrt{-g}=\Lambda V
\end{equation}
where (\ref{CC2}) is  the term  in  the $S_{EH}$. In this case the
notion of conjugation is well-defined, but approximate, as implied
by the expansion about the static Fubini--Study metric (Section
6.1 of \cite{Min1}). Unfortunately, in the proof per se
(\ref{CC1}), relying on the procedure with a non-linear and
non-local Wheeler--de-Witt-like equation of the
background-independent Matrix theory, some unconvincing arguments
are used, making it insufficiently rigorous (Appendix 3 of
\cite{Min1}). But, without doubt, this proof has a significant
result, though failing to clear up the situation.
\\ In\cite{shalyt-aip},\cite{shalyt-entropy2}, \cite{shalyt-gup} the Heisenberg
Uncertainty Relation for the pair $(\Lambda,V)$  (\ref{CC1}) has been generalized to GUP
\begin{equation}\label{CC8}
\Delta V\geq \frac{\hbar}{\Delta
\Lambda}+\alpha_{\Lambda}^{\prime} t_{p}^2
\overline{V}_{p}^{2}\frac{\Delta \Lambda}{ \hbar}
\end{equation}
or that is the same
\begin{equation}\label{CC8.1}
\Delta V\Delta \Lambda \geq \hbar(1+\alpha_{\Lambda}^{\prime}
t_{p}^2 \overline{V}_{p}^{2}\frac{(\Delta
\Lambda)^{2}}{\hbar^{2}}).
\end{equation}
 where $\alpha_{\Lambda}^{\prime}$ is a new constant and $\overline{V}_{p}=l_{p}^{3}.$
\\In the case of UV - limit: $t\rightarrow t_{min}$,$\Delta\Lambda$ becomes significant
\begin{equation}\label{CC10}
\lim\limits_{t\rightarrow
t_{min}}\overline{V}=\overline{V}_{min}\sim
\overline{V}_{p}=l_{p}^{3}; \lim\limits_{t\rightarrow
t_{min}}V=V_{min}\sim V_{p}=l_{p}^{3}t_{p},
\end{equation}
where $\overline{V}$ -- spatial part of ${V}.$
\\ The existence of $V_{min}\sim V_{p}$
directly follows from GUP for the pair $(p,x)$ (\ref{UDP2.1})and
GUP for the pair $(E,t)$ \cite{shalyt3},\cite{shalyt9} as well as
from solutions of the quadratic
inequalities(\ref{CC8}),(\ref{CC8.1}).
\\ So, (\ref{CC8}) is nothing else but
\begin{equation}\label{CC11}
\Delta V \geq \frac{\hbar}{\Delta
\Lambda}+\alpha_{\Lambda}^{\prime} V_{p}^{2}\frac{\Delta
\Lambda}{\hbar}.
\end{equation}
And in the case of UV – cutoff we have
\begin{equation}\label{CC12}
\lim\limits_{t\rightarrow t_{min}}\Lambda \equiv \Lambda_{UV} \sim
\Lambda_{p}\equiv\hbar/V_{p}=E_{p}/\overline{V}_{p}.
\end{equation}
\\It is easily seen that in this case $\Lambda_{UV} \sim m_{p}^{4}$, in
agreement with the value obtained using a standard  (i.e. without
super-symmetry and the like) quantum field theory
\cite{Zel1},\cite{Wein1}. Despite the fact that $\Lambda $ at
Planck's scales (referred to as $\Lambda_{UV} $)  is also a
dynamic quantity, it is not directly related to the familiar
$\Lambda $ because the latter, as opposed to the first one, is
derived from Einstein's equations
\begin{equation}\label{CC13}
R_{\mu \nu} - \frac{1}{2} g_{\mu \nu} R = 8\pi G_N \left( -
\Lambda g_{\mu \nu} + T_{\mu \nu} \right).
\end{equation}
However, Einstein's equations (\ref{CC13}) are not valid at the
Planck scales and hence $\Lambda_{UV}$  may be considered as some
high-energy generalization (deformation) of the conventional
cosmological constant in the low-energy limit.
\\ The problem is whether a correct generalization of GUP for the pair
$(\Lambda,V)$ (\ref{CC8.1}) to the Symmetric Generalized
Uncertainty Principle (SGUP) of the form given by (\ref{SGUP1}) is
possible. If the answer is positive, a theory also includes
$\Lambda_{min}$ that may be referred to as $\Lambda_{IR}$ in
similarity with $\Lambda_{UV}$. Then, similar to Ò (\ref{SGUP1}),
an additional term defining the IR-cutoff must be of the form
\begin{equation}\label{SGUP-cc}
\Omega_{IR}=\frac{(\Delta V)^2}{\widetilde{V}^2},
\end{equation}
where $\widetilde{V}$ - certain space-time volume effectively
specifying the IR-limit of the observable part of the Universe
with the spatial part $\widetilde{\overline{V}}\sim L^{3}$; $L$ --
radius of the observable part of the Universe. Now it is known
that $L\approx 10^{28}$. Clearly, the introduction of an
additional term of the form (\ref{SGUP-cc}) into the right-hand
side of (\ref{CC8.1}) leads to $\Lambda_{IR}\ll \Lambda_{UV}$ and
might lead to the value of $\Lambda$ close to the experimental
value $\Lambda_{exp}$ \cite{Dar1}.
\\ Note that the Holographic Principle \cite{Hooft1}--\cite{Sussk1}
used to the Universe as a whole \cite{Sussk1} gives
$\Lambda_{exp}$  \cite{Bal}. In
\cite{shalyt14},\cite{shalyt-vac},\cite{shalyt-aip},\cite{shalyt-entropy2}
it has been demonstrated that the $\alpha$--representation
($\alpha$--deformation) of QFT with GUP plays a significant role.
In particular, consider
\begin{equation}\label{CC14}
\Lambda_{exp}\approx \alpha_{L} \Lambda_{UV},
\end{equation}
where $L\approx 10^{28}$.(\ref{CC14}) is like  (\ref{EUP4}). But
the Holographic Principle imposes strict restrictions on the
number of degrees of freedom in the Universe, and hence for us it
is important to study the inferences of introducing the additional
term of the form (\ref{SGUP-cc}) in  (\ref{CC8.1}).

\section{GUP, EUP, and General Relativity Deformation}
In this Section we use the previously obtained results for some
cases of high- energy and low-energy deformation of GR.
Specifically, we demonstrate that in the cases when the {\bf
Thermodynamics Approach}
 \cite{Jac1}--\cite{Cai1} is applicable to the General
Relativity the deformation of GR with GUP and EUP may be a natural result of
the $\alpha$-representation.

\subsection{$\alpha$--Representation of Einstein's Equations for
space with horizon}
Let us consider $\alpha$-representation and
high energy $\alpha$-deformation of the Einstein's field equations
for the specific cases of horizon spaces (the point (c) of Section
4). In so doing the results of the survey work (\cite{Padm13}
p.p.41,42)are used. Then, specifically, for a static, spherically
symmetric horizon in space-time described by the metric
\begin{equation}\label{GT9}
ds^2 = -f(r) c^2 dt^2 + f^{-1}(r) dr^2 + r^2 d\Omega^2
\end{equation}
the horizon location will be given by simple zero of the function
$f(r)$, at $r=a$.
\\It is known that for horizon spaces one can introduce
the temperature that can be identified with an analytic
continuation to imaginary time. In the case under consideration
(\cite{Padm13}, eq.(116))
\begin{equation}\label{GT10}
k_BT=\frac{\hbar cf'(a)}{4\pi}.
\end{equation}
Therewith, the condition $f(a)=0$ and $f'(a)\ne 0$ must be
fulfilled.
\\ Then at the horizon $r=a$ Einstein's field equations
\begin{equation}\label{GT11}
\frac{c^4}{G}\left[\frac{1}{ 2} f'(a)a - \frac{1}{2}\right] = 4\pi
P a^2
\end{equation}
may be written as the thermodynamic identity (\cite{Padm13}
formula (119))
\begin{equation}\label{GT12}
   \underbrace{\frac{{{\hbar}} cf'(a)}{4\pi}}_{\displaystyle{k_BT}}
    \ \underbrace{\frac{c^3}{G{{\hbar}}}d\left( \frac{1}{ 4} 4\pi a^2 \right)}_{
    \displaystyle{dS}}
  \ \underbrace{-\ \frac{1}{2}\frac{c^4 da}{G}}_{
    \displaystyle{-dE}}
 = \underbrace{P d \left( \frac{4\pi}{ 3}  a^3 \right)  }_{
    \displaystyle{P\, dV}}
\end{equation}
where $P = T^{r}_{r}$ is the trace of the momentum-energy tensor
and radial pressure. In the last equation $da$ arises in the
infinitesimal consideration of Einstein's equations when studying
two horizons distinguished by this infinitesimal quantity $a$ and
$a+da$ (\cite{Padm13} formula (118)).
\\Now we consider (\ref{GT12}) in a new notation, expressing $a$
in terms of the corresponding deformation parameter $\alpha$.
Hereinafter in this Section we write $\alpha$ instead of $\alpha_{a}$ as we consider the same $a$. Then we
have
\begin{equation}\label{GT13}
a=l_{min}\alpha^{-1/2}.
\end{equation}
Therefore,
\begin{equation}\label{GT14}
f'(a)=-2l^{-1}_{min}\alpha^{3/2}f'(\alpha).
\end{equation}
Substituting this into (\ref{GT11}) or into (\ref{GT12}), we
obtain in the considered case of Einstein's equations in the
"$\alpha$--representation" the following:
\begin{equation}\label{GT16}
\frac{c^{4}}{G}(-\alpha f'(\alpha)-\frac{1}{2})=4\pi
P\alpha^{-1}l^{2}_{min}.
\end{equation}
Multiplying the left- and right-hand sides of the last equation by
$\alpha$, we get
\begin{equation}\label{GT16.1}
\frac{c^{4}}{G}(-\alpha^{2}f'(\alpha)-\frac{1}{2}\alpha)=4\pi
Pl^{2}_{min}.
\end{equation}
But since usually $l_{min}\sim l_{p}$ (that is just the case if
the Generalized Uncertainty Principle (GUP) is satisfied), we have
$l^{2}_{min}\sim l^{2}_{p}=G\hbar/c^{3}$. When selecting a system
of units, where $\hbar=c=1$, we arrive at $l_{min}\sim l_{p}=\surd
G$, and then (\ref{GT16}) is of the form
\begin{equation}\label{GT16.A}
-\alpha^{2}f'(\alpha)-\frac{1}{2}\alpha=4\pi P\vartheta^{2}G^{2},
\end{equation}
where $\vartheta=l_{min}/l_{p}$. L.h.s. of (\ref{GT16.A}) is
dependent on $\alpha$. Because of this, r.h.s. of (\ref{GT16.A})
must be dependent on $\alpha$ as well, i. e. $P=P(\alpha)$.
\begin{center}
{\bf Analysis of $\alpha$-Representation of Einstein's
Equations}
\end{center}
Now let us get back to (\ref{GT12}). In \cite{Padm13} the
low-energy case has been considered, for which (\cite{Padm13} p.42
formula (120))
\begin{equation}\label{GT17.A}
 S=\frac{1}{ 4l_p^2} (4\pi a^2) = \frac{1}{ 4} \frac{A_H}{ l_p^2}; \quad E=\frac{c^4}{ 2G} a
    =\frac{c^4}{G}\left( \frac{A_H}{ 16 \pi}\right)^{1/2},
\end{equation}
where $A_H$ is the horizon area. In our notation (\ref{GT17.A})
may be rewritten as
\begin{equation}\label{GT17.A1}
 S= \frac{1}{4}\pi\alpha^{-1}; \quad E=\frac{c^4}{2G} a
 =\frac{c^4}{G}\left( \frac{A_H}{ 16 \pi}\right)^{1/2}=\frac{\vartheta}{2\surd G}\alpha^{1/2}.
\end{equation}
We proceed to two entirely different cases: low energy (LE) case
and high energy (HE) case. In our notation these are respectively
given by
\begin{center}
A)$\alpha\rightarrow 0$ (LE), B)$\alpha\rightarrow 1/4$ (HE),
\\C)$\alpha$ complies with the familiar scales and energies.
\end{center}
The case of C) is of no particular importance as it may be
considered within the scope of the conventional General
Relativity.
\\Indeed, in point A)$\alpha\rightarrow 0$ is not actually an exact
limit as a real scale of the Universe (Infrared (IR)-cutoff
$l_{max}\approx 10^{28}cm$), and then
\begin{center}
$\alpha_{min}\sim l_{p}^{2}/l^{2}_{max}\approx 10^{-122}$.
\end{center}
In this way A) is replaced by A1)$\alpha\rightarrow \alpha_{min}$.
In any case at low energies the second term in the left-hand side
(\ref{GT16.A}) may be neglected in the  infrared limit.
Consequently, at low energies (\ref{GT16.A}) is written as
\begin{equation}\label{GT16.LE}
-\alpha^{2}f'(\alpha)=4\pi P(\alpha)\vartheta^{2}G^{2}.
\end{equation}
Solution of the corresponding Einstein equation – finding of the
function $f(\alpha)=f[P(\alpha)]$ satisfying(\ref{GT16.LE}). In
this case formulae (\ref{GT17.A}) are valid as at low energies a
semiclassical approximation is true. But from (\ref{GT16.LE})it
follows that
\begin{equation}\label{GT16.solv}
f(\alpha)=-4\pi \vartheta^{2}G^{2}\int
\frac{P(\alpha)}{\alpha^{2}}d\alpha.
\end{equation}
On the contrary, knowing $f(\alpha)$, we can obtain
$P(\alpha)=T^{r}_{r}.$
\\
\begin{center}
{\bf Possible High Energy $\alpha$-Deformation of General
Relativity}
\end{center}
 Let us consider the high-energy case B).
Here two variants are possible.
\begin{center}
{\bf I. First variant}.
\end{center}
In this case it is assumed that in the high-energy (Ultraviolet
(UV))limit the thermodynamic identity (\ref{GT12}) is retained but
now all the quantities involved in this identity become
$\alpha$-deformed. This means that they appear in the
$\alpha$-representation with quantum corrections and are
considered at high values of the parameter $\alpha$, i.e. at
$\alpha$ close to 1/4. In particular, the temperature $T$ from
equation (\ref{GT12}) is changed by $T_{GUP}$ (\ref{UDP2.71}), the
entropy $S$ from the same equation given by semiclassical formula
(\ref{GT17.A}) is changed by $S_{GUP}$ (\ref{UDP2.91}), and so
forth:
\begin{center}
$E\mapsto E_{GUP}, V\mapsto V_{GUP}$.
\end{center}
Then the high-energy $\alpha$-deformation of equation (\ref{GT12})
takes the form
\begin{equation}\label{GT8.GUP}
k_{B}T_{GUP}(\alpha)dS_{GUP}(\alpha)-dE_{GUP}(\alpha)=P(\alpha)dV_{GUP}(\alpha).
\end{equation}
Substituting into (\ref{GT8.GUP}) the corresponding quantities
\\$T_{GUP}(\alpha),S_{GUP}(\alpha),E_{GUP}(\alpha),V_{GUP}(\alpha),P(\alpha)$
and expanding them into a Laurent series in terms of $\alpha$,
close to high values of $\alpha$, specifically close to
$\alpha=1/4$, we can derive a solution for the high energy
$\alpha$-deformation of general relativity (\ref{GT8.GUP}) as a
function of $P(\alpha)$. As this takes place, provided at high
energies the generalization of (\ref{GT12}) to (\ref{GT8.GUP}) is
possible, we can have the high-energy $\alpha$-deformation of the
metric. Actually, as from (\ref{GT12}) it follows that
\begin{equation}\label{GT8.GUP1}
f'(a)=\frac{4\pi k_{B}}{\hbar c}T=4\pi k_{B}T
\end{equation}
(considering that we have assumed $\hbar=c=1$), we get
\begin{equation}\label{GT8.GUP2}
f'_{GUP}(a)=4\pi k_{B}T_{GUP}(\alpha).
\end{equation}
L.h.s. of (\ref{GT8.GUP2}) is directly obtained in the
$\alpha$-representation. This means that, when $f'\sim T$, we have
$f'_{GUP}\sim T_{GUP}$ with the same factor of proportionality. In
this case the function $f_{GUP}$ determining the high-energy
$\alpha$-deformation of the spherically symmetric metric may be in
fact derived by the expansion of $T_{GUP}$, that is known from
(\ref{UDP2.71}), into a Laurent series in terms of  $\alpha$ close
to high values of $\alpha$ (specifically close to $\alpha=1/4$),
and by the subsequent integration.
\\ It might be well to remark on the following.
\\
\\{\bf 4.1.1} As on going to high energies we use (GUP),
$\vartheta$ from equation (\ref{GT16.A})is expressed in terms of
$\alpha^{\prime}$--dimensionless constant from GUP
(\ref{UDP2.1}):$\vartheta=2\alpha^{\prime}.$
\\
\\{\bf 4.1.2} Of course, in all the formulae including $l_{p}$
this quantity must be changed by $G^{1/2}$ and hence $l_{min}$
by $\vartheta G^{1/2}=2\alpha^{\prime} G^{1/2}.$
\\
\\{\bf 4.1.3} As noted in the end of subsection 6.1,
and in this case also knowing all the high-energy deformed
quantities
$T_{GUP}(\alpha),S_{GUP}(\alpha),E_{GUP}(\alpha),V_{GUP}(\alpha)$,
we can find $P(\alpha)$ at $\alpha$ close to 1/4.
\\
\\{\bf 4.1.4} Here it is implicitly understood that the Ultraviolet
limit of Einstein's  equations is independent of the starting
horizon space. This assumption is quite reasonable. Because of
this, we use  the well-known formulae for the modification of
thermodynamics and statistical mechanics of black holes in the
presence of GUP \cite{acs}--\cite{Nou},\cite{Park},\cite{Kim1}.
\\
\\{\bf 4.1.5} The use of the thermodynamic identity
(\ref{GT8.GUP}) for the description of the high energy deformation
in General Relativity implies that on going to the UV-limit of
Einstein's equations for horizon spaces in the thermodynamic
representation (consideration) we are trying to remain within the
scope of {\bf equilibrium statistical mechanics} \cite{Balesku1}
({\bf equilibrium thermodynamics}) \cite{Bazarov}. However, such
an assumption seems to be too strong. But some grounds to think so
may be found as well. Among other things, of interest is the
result from \cite{acs} that GUP may prevent black holes from their
total evaporation. In this case the Planck's remnants of black
holes will be stable, and when they are considered, in some
approximation the {\bf equilibrium thermodynamics} should be
valid. At the same time, by author's opinion these arguments are
rather weak to think that the quantum gravitational effects in
this context have been described only within the scope of {\bf
equilibrium thermodynamics} \cite{Bazarov}.
\\
\\{\bf II. Second variant}.
\\ According to the remark of {\bf 4.1.5},
it is assumed that the interpretation of Einstein's equations as a
thermodynamic identity (\ref{GT12}) is not retained on going to
high energies (UV--limit), i.e. at $\alpha\rightarrow 1/4$, and
the situation is adequately described exclusively by {\bf
non-equilibrium thermodynamics} \cite{Bazarov},\cite{Gyarm}.
Naturally, the question arises: which of the additional terms
introduced in (\ref{GT12}) at high energies may be leading to such
a description?
\\In the \cite{shalyt-gup},\cite{shalyt-aip} it has been shown that in case the cosmological
term $\Lambda$ is a dynamic quantity, it is small at low energies
and may be sufficiently large at high energies.
In the right-hand side of (\ref{GT16.A}) in the $\alpha$--representation the additional term
$GF(\Lambda(\alpha))$ is introduced:
 \begin{equation}\label{GT16.B}
 -\alpha^{2}f'(\alpha)-\frac{1}{2}\alpha=4\pi P\vartheta^{2}G^{2}-GF(\Lambda(\alpha)),
 \end{equation}
 where in terms of $F(\Lambda(\alpha))$ we denote the term including $\Lambda(\alpha)$ as a factor.
  Then its inclusion
in the low-energy case (\ref{GT11})(or in the $\alpha$
-representation (\ref{GT16.A})) has actually no effect on the
thermodynamic identity (\ref{GT12})validity, and consideration
within the scope of equilibrium thermodynamics still holds true.
It is well known that this is not the case at high energies as the
$\Lambda$-term may contribute significantly to make the "process"
non-equilibrium in the end \cite{Bazarov},\cite{Gyarm}.
\\ Is this the only cause for violation of the thermodynamic
identity (\ref{GT12}) as an interpretation of the high-energy
generalization of Einstein's equations? Further investigations are
required to answer this question.
\subsection{$\alpha$--Representation for Friedmann Equations with
GUP and EUP}
Thermodynamic interpretation of Section 4 has been also developed for Friedmann Equations (FEs) of the
Friedmann-Robertson-Walker (FRW) Universe in \cite{Cai1}.
In the process it is taken into consideration that in the FRW space-time, where the metric is
given by the formula
\begin{eqnarray}
ds^2=-dt^2+a^2(\frac{dr^2}{1-kr^2}+r^2d\Omega_{n-1}^2),
\end{eqnarray}
and $d\Omega_{n-1}^2$ denotes a line element of the
($n-1$)-dimensional unit sphere, $a$ is the scale factor, $k$ is
the spatial curvature constant, there is a dynamic  {\bf apparent
horizon}, the radius of which is as follows:
\begin{eqnarray}
\tilde{r}_A=\frac{1}{\sqrt{H^2+k/a^2}},
\end{eqnarray}
where $H\equiv \dot{a}/a$ is the Hubble parameter.
\\ FEs in \cite{Cai1} have been derived proceeding from the assumption that
{\bf apparent horizon} is endowed with the associated entropy and
temperature such the event horizon in the black hole case
\begin{eqnarray}
S=\frac{A}{4G},~~~~~T=\frac{1}{2\pi \tilde{r}_A}
\end{eqnarray}
and from the validity of the first low of thermodynamics
\begin{eqnarray}
dE=TdS.
\end{eqnarray}
\\ In \cite{FRW1} with the use of this thermodynamic interpretation of
FEs the modifications of GUP and EUP (or more precisely the GUP
and EUP deformations) of FEs have been obtained. It is clear that
these (GUP and EUP)--deformed FEs may be written in the form of
the $\alpha$--representation. For simplicity, let us consider the
case $n=3$.
\\ Then for GUP the formula (26) from \cite{FRW1}
takes the form
\begin{eqnarray}
(\dot{H}-\frac{k}{a})[1+\pi\alpha^{\prime 2}l_p^2\frac{1}{A}+
2(\pi \alpha^{\prime 2}l_p^2)^2\frac{1}{A^2}\nonumber\\
+\sum_{d=3}c_d(4\pi\alpha^{\prime 2}
l_p^2)^{2d}\frac{1}{A^d}]=-4\pi G(\rho+p),\label{fr3},
\end{eqnarray}
whereas in the $\alpha$--representation its form is more elegant
\begin{eqnarray}
(\dot{H}-\frac{k}{a})[1+\frac{1}{16}\alpha_{\tilde{r}_A}
+\frac{1}{32}\alpha^{2}_{\tilde{r}_A}
\nonumber\\
+\sum_{d=3}\frac{c_d}{4^{d}}\alpha^{d}_{\tilde{r}_A}]=-4\pi
G(\rho+p),\label{fr31}.
\end{eqnarray}
Also, more elegant is $\alpha$--representation of the
second Friedmann Equation (formula (27) from \cite{FRW1})
\begin{eqnarray}
\frac{8\pi G^{2}}{3}\rho= \frac{\alpha_{\tilde{r}_A}}{4\pi
\alpha^{\prime 2}}[\pi+\frac{1}{32}\alpha_{\tilde{r}_A}
+\frac{1}{96}\alpha^{2}_{\tilde{r}_A}
\nonumber\\
+\sum_{d=3}\frac{c_d}{4^{d}(d+1)}\alpha^{d}_{\tilde{r}_A}],\label{fr4},
\end{eqnarray}
with the assumption that $\hbar=c=1$.
\\ It is obvious that therewith familiar FEs appear at low energies, i.e. at $\alpha_{\tilde{r}_A}\ll 1/4$.
\\ In the nontrivial high-energy case one can obtain the solution for FE
and, in particular $\rho,H,p$ as a series in terms of  $\alpha$ close to 1/4.
\\ In the case of EUP the $\alpha$--representation of the deformed FE \cite{FRW1}
seems to be even simpler. Specifically, using (\ref{EUP1}) --
(\ref{EUP5}), one can derive deformed first Friedmann equation of
the form
\begin{eqnarray}
(\dot{H}-\frac{k}{a^2})(1+\frac{\beta^2}{\pi
l^2}A)=(\dot{H}-\frac{k}{a^2})(1+\frac{\beta^2}{\pi l^2}\frac{4\pi
\tilde{r}_A^{2}}{l^{2}_{or}}l^{2}_{or})\nonumber\\
=(\dot{H}-\frac{k}{a^2})(1+4\beta^2\alpha^{-1}_{\tilde{r}_A}\alpha_{l_{or}})
=(\dot{H}-\frac{k}{a^2})(1+4\beta^2\widetilde{\alpha}_{\tilde{r}_A})=-4\pi
G(\rho+p),
\end{eqnarray}
where, as expected, the deformation parameter   $\widetilde{\alpha}_{\tilde{r}_A}$ is small.
\\ In a similar way we can obtain the $\alpha$--representation of the
EUP-deformation for the second Friedmann equation.
\section{Some Comments and Problems of Interest}
In this Section some comments are given and some problems are stated. \\
\\Ñ1. The Laurent series expansion in terms of $\alpha$ is asymmetric  for UV and IR cutoffs.
Indeed, as in the general case the variability domain
$0<\alpha\leq1/4$, in the UV-cutoff when $\alpha\approx 1/4$ the
contribution is made by $\alpha$-terms both with positive and with
negative powers, while in the IR-cutoff ($\alpha\ll 1/4$) only the
$\alpha$-terms with negative powers will be significant.
\\
\\C2. The external constant $\alpha^{\prime}$ in the cases,
where $l_{or}\neq l_{min}$ (EUP or SGUP is the case), is not found
in the final expressions, being reduced due to the substitution
of(\ref{EUP4}).
\\
\\ Several questions remain to be answered and necessitate further investigations.
\\
\\Q1) How far the $\alpha$-representation may be extended for the
General Relativity? As shown in this work, such a representation
exists for the General Relativity at High and Low Energies when
the {\bf Thermodynamic Approach} \cite{Jac1}--\cite{Cai1} is
applicable or, that is the same, the Thermodynamic Interpretation
 is the case. It is interesting whether the extension of
the $\alpha$-representation to the general case both at High and
Low Energies is possible. The problem is whether, in some or other
way, the general case may be reduced to the well-known ones.
\\
\\Q2) Considering Q1), for High Energy the problem is whether
there is an effective description of the space-time foam
\cite{Wheel}--\cite{Isham} in terms of $\alpha$. The results of
\cite{Gar2} suggest that such a description should be existent.
\\
\\Q3) Proceeding from the results of E.Verlinde \cite{Verl},
the problem is whether the High-Energy deformation of the
Entropic Force  is obtainable. Provided the answer of Q1)
positive, the problem concerns the form of this deformation in
terms of $\alpha$: we must find its $alpha$-representation.
\\Note that the notion of  Entropic
Force, however, without the introduction of the term per se has
been proposed  by T.Padmanabhan in Conclusion of his paper
\cite{Padm-new} earlier than by E.Verlinde.

\section{Conclusion}
In the case the problems stated in the previous Section will be
solved positively, the small dimensionless discrete parameter
$\alpha$ must be at once introduced in Íigh-Energy Thermodynamics
and Gravity, without its appearance in the low-energy limit at the
scales under study. At the same time, at large scales GR has not
been subjected to verification too \cite{Tur}. The availability of
Dark Matter and Dark Energy is a strong motivation for the
IR-modification of GR \cite{Dvali}-- \cite{Rub}. The deformation
of the General Relativity due to EUP seems to be one of the
IR-modifications of Gravity possible. In this case an analysis of
such a deformation in terms of the parameter $\alpha$, of the
corresponding variability domain, and the like may be important
for studies of the IR-modified (IR-deformed) General Relativity.

\section{Acknowledgments}
I am grateful to Prof. Sabine Hossenfelder (Stockholm, Sweden) for
the information about a number of interesting works in the field
of physics at Planck's scales and for her support that has
stimulated my research efforts. Besides, I would like to thank
Prof. Rong-Jia Yang (Baoding, China) for his support.

\end{document}